\documentclass{owrart}

\newcommand{\eps}{\epsilon}
\newcommand{\IR}{\mathbb{R}}
\newcommand{\cO}{\mathcal{O}}
\newcommand{\cP}{\mathcal{P}}
\newcommand{\cL}{\mathcal{L}}
\theoremstyle{plain}
\newtheorem{thm}{Theorem}

\begin{document}


\begin{talk}{St{\'e}phane Nonnenmacher}
{Quantum transfer operators and chaotic scattering}
{Nonnenmacher, Stephane}

\noindent

Consider a symplectic diffeomorphism $T$ on 
$T^*\IR^d$, which can be generated near the origin by a single function $W(x_1,\xi_0)$, in the sense that the dynamics 
$(x_1,\xi_1)=T(x_0,\xi_0)$ is the implicit 
solution of the two equations $\xi_1=\partial_{x_1}W(x_1,\xi_0)$, $x_0=\partial_{\xi_0}W(x_1,\xi_0)$. 
One can associate to $T$ 
a family of {\it quantum transfer operators}  $M(T,h)$
acting on $L^2(\IR^d)$, of the form:
\begin{equation}\label{FIO}
[M(T,h)\psi](x_1)=\int a(x_1,\xi_0)\,e^{\frac{i}{h}(W(x_1,\xi_0)-\langle \xi_0,x_0\rangle)}\,\psi(x_0)\,
\frac{dx_0\,d\xi_0}{(2\pi h)^d}\,.
\end{equation} 
Here $a\in C^\infty(T^*\IR^d)$ is called the {\it symbol} of the operator. The ``small parameter'' $h>0$ 
is the typical 
wavelength on which the integral kernel of the operator oscillates; 
it is often called ``Planck's constant'', due to the 
appearance of such operators in quantum mechanics. 

The operator $M(T,h)$ (understood as a family $(M(T,h))_{h\in (0,1]}$) can be interpreted 
as a ``quantization'' of the symplectic map $T$, for the following reason. 
Consider a phase space point $(x_0,\xi_0)\in T^*\IR^d$. There exist 
wavefunctions ({\it quantum states}) $\psi_{x_0,\xi_0,h}\in L^2(\IR^d)$ which are localized near the position $x_0\in \IR^d$, and whose
$h$-Fourier transform is localized near the momentum $\xi_0\in\mathbb{R}^d$ (equivalently, the usual Fourier transform is localized
near $h^{-1}\xi_0$). Such wavefunctions are said to be {\it microlocalized} near $(x_0,\xi_0)$; in some sense, 
they represents the best quantum approximation of a ``point particle'' at $(x_0,\xi_0)$.
In the semiclassical limit $h\to 0$, the application of stationary phase expansions to the integral (\eqref{FIO}) 
shows that the image state 
$M(T,h)\psi_{x_0,\xi_0,h}$ is microlocalized near the point $(x_1,\xi_1)=T(x_0,\xi_0)$; that is, this operator transports the
quantum mass at the point $(x_0,\xi_0)$ to the point $T(x_0,\xi_0)$.

Similar families of operators have appeared in the theory of linear PDEs
in the 1960s: the ``Fourier integral operators'' invented by H\"ormander. 
A modern account (closer to the above definition) 
can be found in the recent lecture notes of C.Evans \& M.Zworski \cite{sn_EV}. 
We are using these
operators as nice models for 
``quantum chaos'', that is the study of quantum systems, the classical limits of which are ``chaotic''. 
In this framework, these operators (sometimes called ``quantum maps'') generate a quantum dynamical system: 
\begin{equation}\label{DS}
L^2(\IR^d)\ni\psi\mapsto M(T,h)\psi\,.
\end{equation}
These quantum maps provide a discrete time generalization of the Schr\"odinger flow $U^t(h)=\exp(-itP(h)/h)$ associated with 
the Schr\"odinger equation 
$ih\partial_t \psi=P(h)\,\psi$, where $P(h)$ is a selfadjoint operator, e.g. of
the form $P(h)=-\frac{h^2\Delta}{2} +V(x)$; in that case, the classical evolution is the Hamilton flow $\phi_p^t$ generated by the 
classical Hamiltonian $p(x,\xi)=\frac{|\xi|^2}{2}+V(x)$ on $T^*\IR^d$. 

As usual in dynamics, one is mostly interested in the long time properties of the dynamical system \eqref{DS}. For such a 
linear dynamics, these properties are encoded in the {\it spectrum} of $M(T,h)$. Therefore, a major focus of investigation concerns
the spectral properties of the operators $M(T,h)$, especially in the semiclassical limit $h\to 0$, where the connection to
the classical map is most effective. Quantum maps have mostly be studied in cases where $M(T,h)$ is replaced by a unitary operator
on some $N$-dimensional Hilbert space, with $N\sim h^{-1}$. This is the case if $T$ is a symplectomorphism on a compact 
symplectic manifold, like the 2-torus \cite{sn_DEG03}.
More recently, one has got interested in operators $M(T,h)$ which act unitarily
on states microlocalized 
inside a certain domain of $T^*\IR^d$, but ``semiclassically kill'' states
microlocalized outside a larger bounded domain (these properties
depend on the choice of the symbol $a(x_1,\xi_0)$).
As a result, the spectrum of $M(T,h)$ is contained in the unit disk, and its effective rank is $\leq C h^{-d}$ 
(according to the handwaving argument that one quantum state
occupies a volume $\sim h^d$ in phase space). Such operators have been called ``open quantum maps''.

Let us now assume that the map $T$ has chaotic properties: the nonwandering set $\Gamma$ is
a fractal set included inside  $B(0,R)$, and $T$ is uniformly hyperbolic on $\Gamma$. We may then expect
this dynamical structure to imply some form of {\it quantum decay}: indeed, a quantum state cannot be localized on a ball of radius
smaller than $\sqrt{h}$, and such a ball is not fully contained in $\Gamma$, so most of 
the ball will escape to infinity through the map $T$. On the other hand, quantum mechanics involves {\it interefence effects}, which may balance
this purely classical decay. Following old works of M.Ikawa \cite{sn_Ik} and P.Gaspard \& S.Rice \cite{sn_GaRi} in the framework of Euclidean
obstacle scattering, one is lead to the following condition for quantum decay:
\begin{thm} For any $(x,\xi)\in\Gamma$, call $\varphi^u(x,\xi)=-\log|\det DT_{| E^u(x,\xi)}|$ the unstable Jacobian of $T$ at $(x,\xi)$, and consider
the corresponding topological pressure $\cP(\frac12\varphi^u)$.

If that pressure is {\it negative}, then for any $1>\gamma>\exp\{\cP(\frac12\varphi^u)\}$, and any small enough $h>0$, 
the operator $M(T,h)$ has a spectral radius $\leq \gamma$.
\end{thm}
In dimension $d=1$ (that is, when $T$ acts on $T^*R$), the negativity of that pressure is equivalent with the fact that the Hausdorff
dimension $d_H(\Gamma)<1$. This equivalence breaks down in higher dimension, but a negative pressure is still
correlated with $\Gamma$ being a ``thin'' set. 

The above theorem has been obtained by M.Zworski and myself 
in the framework of Euclidean scattering by smooth potentials \cite{sn_noz-acta09}. 
The extension to quantum maps $M(T,h)$ is straightforward, and should
be part of a work in preparation with J.Sj\"ostrand and M.Zworski. In general we do not expect the above to be optimal.
Following a recent work of V.Petkov \& L.Stoyanov \cite{sn_PetStoy}, 
one should be able to compare $M(T,h)$ with classical transfer operators of the form 
$\cL_{\frac12\varphi^u+i/h}$, apply
Dolgopyat's method to the latter to get a spectral radius
$\gamma = \exp\{\cP(\frac12\varphi^u)-\epsilon_1\}$ for the classical and the quantum operators.

Most of the $\cO(h^{-d})$ eigenvalues of $M(T,h)$ can be very close to the origin when $h\to 0$. Indeed, 
the fractal character of the trapped set has a strong influence on the semiclassical density of
eigenvalues: any point situated at distance $\gg h^{1/2}$ from $\Gamma$ will be pushed out of $B(0,R_1)$ through the classical
dynamics  (either in the past or in the future), within a time $|n|\leq C\log(1/h)$, where semiclassical methods still apply. 
As a result, the eigenstates
of $M(T,h)$ associated with nonnegligible eigenvalues must be ``supported'' by the tubular neighbourhood
of $\Gamma$ of radius $\sqrt{h}$. A direct volume estimate of this neighbourhood, and the above-mentioned 
argument on the volume occupied by a quantum states, lead to the following upper bound for the density of eigenvalues:
\begin{thm}
Assume that the hyperbolic trapped set $\Gamma\subset T^*\IR^d$ has upper Minkowski dimension $d_M>0$. Then, for any small $\eps,\eps'>0$ and
any small enough $h>0$, one has
\begin{equation}\label{fractal}
\# \{\lambda\in \mathrm{Spec}(M(T,h)),\,|\lambda|>\eps\}\leq C_{\eps,\eps'}\,h^{-d_M/2-\eps'}\,.
\end{equation}
(eigenvalues are counted with multiplicities.)
\end{thm}
A similar result has been first obtained by J.Sj\"ostrand in the case of Euclidean scattering by 
smooth potentials \cite{sn_SjDuke}, and has then been
refined and generalized to various settings. The case of quantum maps should also appear in the forthcoming work with J.Sj\"ostrand and M.Zworski. 

The ``fractal upper bound'' (\eqref{fractal}) is actually conjectured to be an asymptotics.
This has been shown numerically in various cases, including hyperbolic scattering \cite{sn_GLZ} as well as quantum maps \cite{sn_SchoTwo}.
This asymptotics has been proved only for a very specific quantum maps \cite{sn_noz-cmp07}, and represents an interesting challenge for more
realistic systems.

\end{talk}

\end{document}